\documentclass[prl,aps,twocolumn,floatfix]{revtex4}
\usepackage{latexsym,amssymb}
\usepackage{graphicx}

\newcommand{\pderiv}[2]{\frac{\partial #1}{\partial #2}}

\begin{document}



\title{Single Polymer Confinement in a Slit: Correlation between structure and dynamics}

\author{Joshua Kalb and Bulbul Chakraborty}

\affiliation{Martin Fisher School of Physics, Brandeis University,
Mailstop 057, Waltham, Massachusetts 02454-9110, USA}
\date{\today}

\begin{abstract}
In this paper, we construct an effective model for the
dynamics of an excluded-volume chain under confinement, by extending the formalism of Rouse modes.
We make specific predictions about the behavior of the modes for a single polymer confined to a slit.
The results are tested against Monte Carlo
simulations using the bond-fluctuation algorithm which uses a
lattice representation of the polymer chain with excluded-volume
effects.\end{abstract}


\maketitle

\section{Introduction}

Confining polymers inevitably leads to a competition between length scales arising from internal cooperativity and those
imposed by external geometry, resulting in qualitative changes of the static and dynamic behavior.  The effects of
confinement on a single polymer chain is still not completely understood. A scaling picture has been used to describe the changes
in the relaxational dynamics of a  single, self-avoiding  chain confined to a tube or a pore\cite{Daou82,Broc77}.
Since a  self-avoiding chain under cylindrical confinement models a variety of phenomena ranging from DNA
translocation through synthetic pores\cite{Aksi04} to bacterial chromosome segregation\cite{Lewi01}, there have been a
number of simulation studies investigating its statics and dynamics\cite{Daou82,Gres88,Thir05,Jend03,Hagi98,Suck07}.
In this paper, we use theory and simulations to study the dynamics of a single,
self-avoiding polymer confined to a slit.

One of the key concepts in polymer dynamics is the relaxation of modes that describe the internal dynamics of a
chain molecule\cite{Doi95,Broc77,Khok93}.  We construct a theoretical framework for
describing the relaxation of the modes of a
confined, self-avoiding chain in the absence of hydrodynamic interactions, by extending the Rouse model\cite{Doi95}.  Our
analysis shows that the anisotropy, introduced by the slit geometry, leads to dramatically different relaxations  of
the longitudinal and transverse modes.   This effect, in turn, changes the nature of the motion of single monomers,  and
the collective dynamics of the self-avoiding chain.  We perform simulations using the bond-fluctuation model\cite{Krem88}, and show that
the results  can be explained, semi-quantitatively, by our theoretical model.   Explicit theoretical
results for the relaxation of Rouse modes, presented this work, provide a powerful tool for predicting fine
scale and large scale motion of biopolymers within the confines of a cell.

\section{Rouse Modes}
If excluded volume and hydrodynamic interactions are
ignored, the dynamics of an unconfined,  ideal chain (without self-avoidance)  is well
described by the Rouse model\cite{Doi88,Doi95}:
\begin{equation}
\zeta\pderiv{{\bf x}_n}{t}=k({\bf x}_{n+1}-2{\bf x}_n+{\bf
x}_{n-1})+{\bf g}_{n}{\bf (t)} \label{eqn:discretelang2}
\end{equation}
where ${{\bf x}_n}$ denotes the position of the $n^{th}$ monomer of size $b$, $k={k_BT \over {b^2}}$,  ${\bf g}_{n}(t)$
is a random force arising from the solvent surrounding the polymer, and $\zeta$ is the friction
constant of a single monomer.
Considering dynamics at length scales much larger than the individual monomer size $b$,
\begin{equation}
{\bf x}_{n+1}-2{\bf x}_n+{\bf x}_{n-1}\approx{\partial^2 {\bf x}(n)\over \partial n^2}  ~,\label{differenceisderivative}
\end{equation}
 and Eq. (\ref{eqn:discretelang2}) becomes a partial differential equation:
\begin{equation}
\zeta\pderiv{{\bf x}(n,t)}{t}=k\frac{\partial^2 {\bf x}(n,t)}{\partial n^2}+{\bf g}(n,t) ~.
\end{equation}
The random force, ${\bf g}{(n,t)}$, is Gaussian and is characterized by the following correlations:
\begin{equation}
<{g}_j{(n,t)}>={\bf 0}
\end{equation}
\begin{equation}
<{g}_j(n,t){g}_k{(n',t')}>={{{2k_B}T\zeta}
\over{N}}\delta_{jk}\delta({t-t'})\delta({n-n'})
\end{equation}
with $j$ and $k$ being the cartesian components of $\bf{g}(n,t)$, and $N$ the number of monomers in the chain.
The normal modes of the above  equation are the Rouse modes, and  they decay exponentially
with characteristic time scales that depend on the mode
index\cite{{Doi88,Doi95}}.

A major missing piece in the above description is excluded volume
interactions\cite{Doi88}. The inclusion of these interactions (or confinement)
introduces couplings between the Rouse
modes, and is described by an equation of the form:
\begin{equation}
\zeta\pderiv{{\bf x}(n,t)}{t}=k\frac{\partial^2 {\bf x}(n,t)}{\partial n^2}+{\bf g}({\bf x},t)+{\bf F}(n,t)
\end{equation}
where ${\bf F}(n,t)$ represents the force on the $n^{th}$ monomer due to all the other monomers, or due to the
confining geometry.  The presence of this term couples the Rouse modes, and in general, this equation has no analytic
solution. An approach that has been used for unconfined, self-avoiding, polymers is to assume
that the dynamics can still be described by decoupled Rouse
modes\cite{Doi88}, but with renormalized parameters. Bond
fluctuation algorithm simulations indicate that for such
polymers, the Rouse modes are indeed weakly
coupled\cite{Krem88} and that the renormalized Rouse modes describe
the dynamical behavior.  The relaxation of the modes is characterized by a time scale $\tau_{p} \simeq {N^{2\nu} \over
p^{2\nu +1}}$, where the mode index $p= 0,1,2, .....$ , and $\nu$ is the Flory exponent: $\nu = 3/(d+2)$ with $d$ being
the spatial dimension\cite{Doi95}. In 2D, $\tau_{p} \simeq {N^{3/2} \over p^{5/2}}$.
The
longest relaxation time is $\tau_{1} \propto N^{2\nu +1}$, and is associated with the reorientation or rotation of the
polymer\cite{Doi95}.

\subsection{Relaxation of Rouse Modes for Confined Polymers}
In this work, we construct a framework for describing Rouse modes in confined polymers. To facilitate discussion, we
briefly outline the steps leading to the results for unconfined polymers with self-avoidance. Neglecting
the coupling between  Rouse modes leads to a set of  linear
equations for the mode amplitudes $X_{pj}$\cite{Doi95,Moli06}:
\begin{equation}
\zeta_{p}\pderiv{X_{pj}}{t}=-k_{pj}X_{pj}+g_{pj}{(t)}~.
\label{eqn:linrouse}
\end{equation}
Here
\begin{equation}
X_{pj}={1\over N} \int_0^N{dn \cos\left(p\pi n \over N\right)x_{nj}(t)} ~,\label{eqn:modeeq}
\end{equation}
and $\zeta_{p} =  2 N \zeta$ for $p > 0$, $\zeta_{0} = N \zeta$.
All the self-avoidance and boundary effects are incorporated through
the $k_{pj}$ term. Requiring that in the long time limit, the Rouse
modes have the correct equilibrium distribution, implies
that\cite{Doi88}:
\begin{equation}
k_{pj}={k_B T \over <X_{pj}^2>_{eq}} \label{eqn:equipart}
\end{equation}
where $<.>_{eq}$ is an equilibrium average. The exact expression for $<X_{pj}^2>_{eq}$ can be derived using Eq.(\ref{eqn:modeeq}):
\begin{eqnarray}
<X_{pj}^2>_{eq}&=&-{1\over 2N^2}\int_0^N\int_0^N dndm \cos\left({p\pi n\over N}\right)\nonumber\\
 & &\cos\left({p\pi m\over N}\right)\times<(x_{nj}-x_{mj})^2>
\label{eqn:modeeqexact}
\end{eqnarray}
As shown by Doi and Edwards\cite{Doi88}, for $p\gg 1$, $<X_{pj}^2>_{eq}$ can be expressed as:
\begin{eqnarray}
<X_{pj}^2>_{eq} & & = {N \over {4 p^2 \pi^2}} \int_{0}^{N}{du}\cos\left(p\pi u \over N\right)\times\nonumber\\
                & & \frac{\partial^2<(x_{mj}-x_{nj})^2>_{eq}}{\partial n
\partial m} \label{eqn:modeeqapprox}
\end{eqnarray}
where $u=m-n$.

The generalized Rouse model treats a confined, self-avoiding polymer as an  ideal polymer with renormalized spring
constants. The utility of the approach relies on access to reliable results for the static correlations between
monomers, $<(x_{mj}-x_{nj})^2>_{eq}$.   The time scales associated with the relaxation of the Rouse modes, and ultimately the dynamics of the
chain and single monomers can be obtained, once these static correlations are known.
The task at hand is, therefore, to find an appropriate model for the equilibrium correlations for confined polymers and
to predict the dynamical behavior using Eqs. (\ref{eqn:linrouse})-(\ref{eqn:modeeqapprox})\cite{Arno07}.

The next section is devoted to discussing appropriate forms for static correlations function of a confined, self-avoiding polymer.

\section{Blob Model for Slit Geometry}
For an unconfined polymer, consisting of $N$ monomers of size $b$, represented as a self-avoiding-walk (SAW),
the radius of gyration scales as\cite{Flor48}:
\begin{eqnarray}
R_{g}&\sim &bN^{\nu}\nonumber \\
{\nu}&=&{3 \over {d+2}}
\end{eqnarray}
 \begin{figure}[h]
 \centering
 \includegraphics[width=\columnwidth]{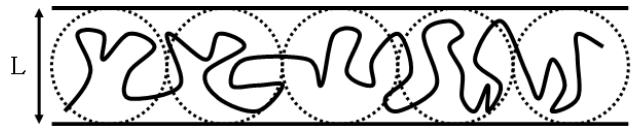}
 \caption{Sample configuration from simulation of a 2-D polymer confined to a tube.
 Circles indicate rough shape of the blobs and have a size on the order of the tube width, $L$.}
 \label{blobpicture}
 \end{figure}
Upon confining this SAW to a slit, we expect the excluded volume
effects to change the monomer-monomer correlation functions.
The scaling of these correlations with the tube width and length of the polymer has
often been described by the blob model\cite{Broc77}. Mean-field approaches have
also been used to model these static correlations\cite{Thir05,Edwa79}.

In a slit of width $L$, the first effect of confinement is
to make the polymer conformations anisotropic\cite{Vlie93}, and the shape needs to be characterized by a transverse
($R_{\perp}$) and a longitudinal ($R_{||}$) length scale.
For widths, $L<R_{g}$, where $R_g$ is the radius of gyration of the
unconfined polymer, confinement has the effect of stretching the
polymer in the axial direction and screening out the excluded volume
interactions beyond a length scale comparable to $L$. The polymer
can be visualized as a succession of blobs with excluded-volume
effects being maintained for distances smaller than the blob
size\cite{Broc77}. The number of monomers per blob, $g$, can be
calculated by requiring that the radius of gyration of these
monomers matches the blob length scale\cite{Azum99,Saka06}.
For the
slit geometry, the length scale of  the blob is assumed to be $L$\cite{Arno07}, and
therefore,
\begin{eqnarray}
g \sim \left({L \over b}\right)^{1/\nu} \label{eqn:mperblob}
\end{eqnarray}
The polymer is composed of $N_b = N/g$ blobs and has a lateral extension
of:
\begin{equation}
R_{||}{=}LN_b = N b (b/L)^{1/\nu -1 } = N b (b/L)^{1/3} ~, \label{eqn:blobperlength}
\end{equation}
where the last relation is specific for the slit geometry (2D).
Since the tube width confines the polymer in the
perpendicular direction\cite{Pinc76}:
\begin{equation}
R_{\perp}{\propto}L
\end{equation}

In papers by Edwards and Singh, estimates of
polymer sizes were made using a mean-field approach\cite{Edwa79}. This method
was later applied to a polymer in a tube by Morrison and
Thirumalai\cite{Thir05}. The blob predictions for the size of a
polymer in narrow tubes agree with their estimates up to a factor of
order unity.

\subsection{Monomer-Monomer correlations in the Blob Model}
The arguments presented above hold equally well for  sub-chains as long as the sub-chains we consider are large enough to have the asymptotic behavior of a SAW.
Therefore,
the correlations between monomers in the directions parallel and perpendicular to the slit walls are should scale as:
\begin{equation}
<(x_{n||}-x_{m||})^2> \sim \left\{ \begin{array}{ll}
         |n-m|^{3/2}b^2 & \mbox{$|n-m| < \left(L \over b\right)^{4/3}$}\\
        |n-m|^2b^{8/3}L^{-2/3} & \mbox{$|n-m| > \left(L \over b\right)^{4/3}$}\end{array} \right .\
\label{eqn:blobscalingpar}
\end{equation}
\begin{equation}
<(x_{n\perp}-x_{m\perp})^2> \sim \left\{ \begin{array}{ll}
         |n-m|^{3/2}b^2 & \mbox{$|n-m| < \left(L \over b\right)^{4/3}$}\\
        L^2 & \mbox{$|n-m| > \left(L \over b\right)^{4/3}$} \end{array} \right .\
\label{eqn:blobscalingperp}
\end{equation}

Recent simulations\cite{Arno07} have measured the distribution of blob sizes by analyzing sub chains, and for strong
confinement ($L < R_{g}$), their results are in agreement with the blob model predictions. In order to proceed with the
dynamical calculations,
we assume that the correlations are piecewise continuous with sharp transitions  between the different scaling
regimes represented in Eqs. (\ref{eqn:blobscalingpar}) and (\ref{eqn:blobscalingperp}).

\subsection{Simulation results for the Monomer-Monomer correlation function}
We have used simulations to measure the monomer correlations and compared them to the predictions of the blob model
outlined above. To our knowledge, these predictions have never been directly confirmed.
\paragraph{Simulation Technique}
\begin{figure}
 \centering
 \includegraphics[width=\columnwidth,clip=true]{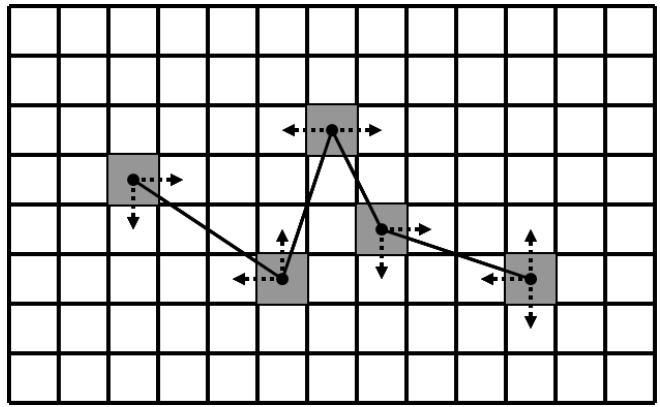}
 \caption{Example chain configuration. Monomers occupy each box.
 Allowed moves are labeled with dashed arrows. Bonds are represented by solid lines.}
 \label{allowedmoves}
 \end{figure}
 The simulations were based on the bond fluctuation algorithm, a
course-grained lattice-based algorithm that allows for the analysis
of dynamical properties of polymers\cite{Krem88}. As the name
implies, the algorithm allows for fluctuations of bond lengths and
employs only local moves of monomers. Excluded-volume effects are
taken into account by ensuring the initial polymer configuration
contains no intersecting bonds and by allowing only the following
set of bonds:
\begin{eqnarray}
S &=& (2,0)\cup(2,1)\cup(2,2)\cup(3,0)\cup(3,1)\cup(3,2)\nonumber
\end{eqnarray}
This also leads to an effective monomer size $b$, where $b\approx
2.8$ lattice spacings. This algorithm has been shown to reproduce
the right scaling for polymer size and follow the generalized Rouse
dynamics for excluded-volume polymers\cite{Doi95} in all spatial
dimensions. It has been widely used to study the dynamics in
different environments\cite{Cecc02,Chen00,Gilr01,Azum99}. In this paper,
the simulation results are used to test the predictions of our  theoretical formalism, and to explore in detail the
dynamics of a chain confined to a slit.

\paragraph{Results}Monomer correlations were obtained for chains  with $N=200$, $R_{g}= 149$. This size is optimal in the
sense that it is large enough to illustrate the effects of blobs while still yielding  relaxation times that can be
accommodated within accessible simulation times.  As will be shown from our calculations of the Rouse mode relaxations,
the longest time scale increases as $N^{3}$ and as $L^{-2/3}$, making it difficult to determine correlations functions
for long chains in a narrow slit.

\begin{figure}[htb]
\centering
\includegraphics[width=\columnwidth,clip=true]{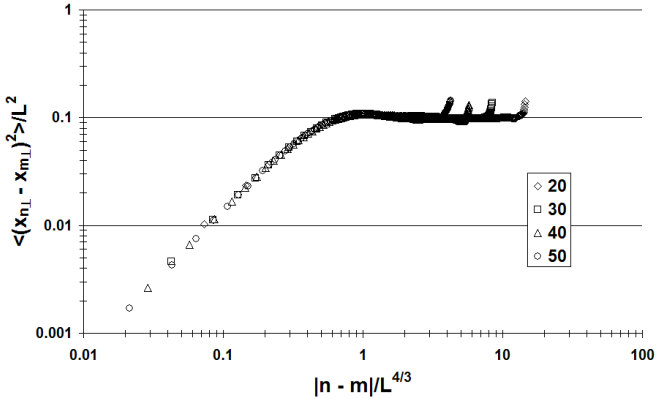}
\caption{Scaling of $<(x_{n\perp}-x_{m\perp})^2>$ , demonstrating the the scaling of $g(L)$ as $L^{4/3}$, and validity of Eq. \ref {eqn:blobscalingperp}. The x-axis has been scaled by
$L^{4/3}$, and th y-axis has been scaled by $L^{2}$.
The chain length is $N=200$, $R_{g}=149$, and the slit widths are 20, 30, 40, and 50.  All lengths are measured in
units of the lattice spacing}
\label{xnxmperpscaled}
 \end{figure}

\begin{figure}[htb]
 \centering
\includegraphics[width=\columnwidth,clip=true]{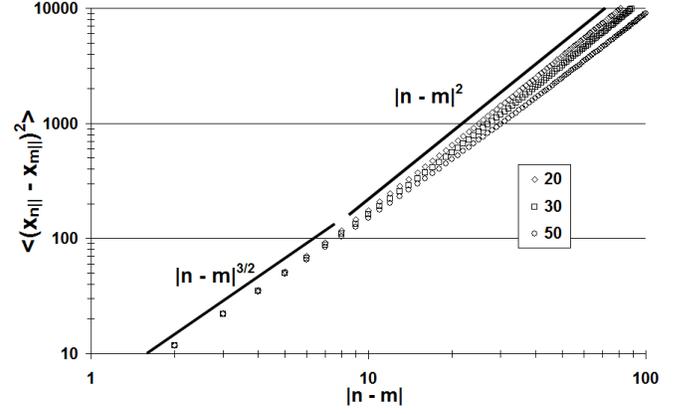}
 \caption{$<(x_{n||}-x_{m||})^2>$ versus slit width for the same set of parameters as Fig. \ref{xnxmperpscaled}.  A crossover between two power laws is observed.}
 \label{xnxmpar}
 \end{figure}

The initial chains were generated using a back-tracking algorithm which would redraw
the part of the walk if a new segment had not been placed within $30$ attempts. To ensure
little biasing from the generation of the initial walk, chains were equilibrated by allowing the
chain to undergo $10^8$ monte carlo steps (mcs) ($200*10^8$ attempted moves). After equilibration,
data was sampled every $10^3$ mcs for runs of $10^8$ mcs. The process was repeated for $10$ initial configurations and averaging was
conducted over all sampled configurations of all runs as well as over the whole chain. This averaging leads to the data shown in Figs. \ref{xnxmperpscaled} and \ref{xnxmpar}.

The scaling laws observed in simulations are in broad agreement with
the blob picture for the monomer correlations perpendicular to (Fig.\ref{xnxmperpscaled}) and parallel to
(Fig.\ref{xnxmpar}) the walls.  In reporting the results of the simulations, we  measure all lengths in units of $b$.
Fig. \ref{xnxmperpscaled} demonstrates that the data for the transverse correlations for different values of $L$ can be
made to scale if $|n-m|$  is divided by $L^{4/3}$ and the correlations are scaled by $L^{2}$, confirming the form of Eq.
\ref{eqn:blobscalingperp}, and the scaling of the blob size $g(L)$ as $L^{1/\nu}$ ($\nu=3/4$ in 2D).
The results for the correlations in the longitudinal direction indicate  a crossover between two power laws, as
shown in Fig. \ref{xnxmpar}.

According to the authors' knowledge, this is
the first direct test through simulations of distinct scaling regimes predicted by the blob model. Since there is broad support for the blob-scaling picture of monomer correlations from the
simulations, we proceed to utilize these results for calculating the properties of the Rouse modes under confinement.

\begin{figure}[htb]
 \centering
 \includegraphics[width=\columnwidth,clip=true]{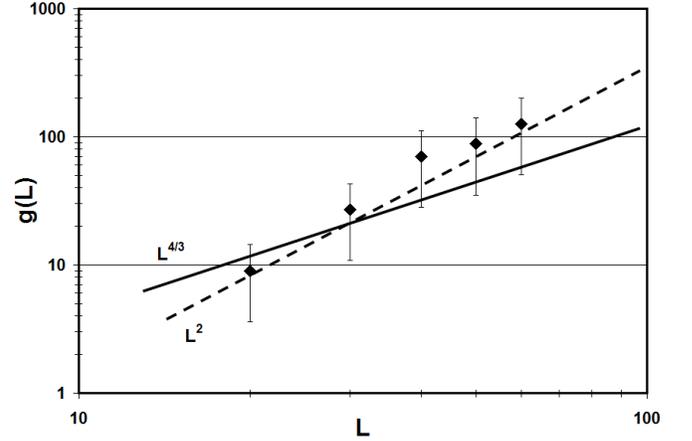}
 \caption{Scaling of the blob size, $g(L)$, extracted from the crossover points of $<(x_{n||}-x_{m||})^2>$ shown in  Fig.\ref{xnxmpar}.  The best fit is for $L^2$, however, the $L^{4/3}$ behavior evident in the transverse correlations is within the error bars on this data.}
 \label{gvsLpar}
\end{figure}

\section{Relaxations of Rouse Modes}
To study the dynamical correlations in the anisotropic geometry of the slit, we analyze the transverse modes, $X_{p\perp}$, and longitudinal modes , $X_{p||}$, separately.
\paragraph{Transverse Modes} The transverse modes correspond to taking $j$ to be in the direction perpendicular to the
slit axis.
The blob picture provides a natural scale of $p_{max}= N/g = N b^{4/3}/L^{4/3}$,
separating the large $p$ from the small $p$ regime.
For $p < <  p_{max}$, we use Eq.(\ref{eqn:blobscalingperp}) directly in Eq.(\ref{eqn:modeeqexact}).   For these modes,
$<X_{p\perp}^2>_{eq} \simeq A - B (p/p_{max})^{2}$, where the constant $A = {L^{10/3} \over {8 \pi N b^{4/3}}}$.
In the regime where $L << R_{g}$,  modes with $p > p_{max}$, satisfy the condition for the use of Eq.  (\ref {eqn:modeeqapprox}), since these
mode numbers satisfy $p > N (b/R_{g})^{4/3} = N$.  Combining Eqs. (\ref{eqn:blobscalingperp}) and (\ref{eqn:modeeqapprox}) we, therefore obtain, for $p>p_{max}$,
\begin{equation}
<X_{p\perp}^2>_{eq} = {N\over 4p^2\pi^2}
\int_{0}^{(L/b)^{4/3}}{du}\cos\left(p\pi u \over N\right){b^2 \over
u^{1/2}} \label{eqn:xperp}
\end{equation}
\begin{equation}
<X_{p\perp}^2>_{eq} = {N^{3/2}b^2 \over
2^{3/2}p^{5/2}\pi^2}F_c\left(\sqrt{2 p \over p_{max}}
\right) \label{eqn:Xperp}
\end{equation}

In the above equation, and below, we set $b=1$, and $\zeta /{k_{B} T} = 1$.
The
relaxation of the transverse modes is exponential with a time scale:
\begin{equation}
\tau_{p\perp}={\zeta_p \over k_{p\perp}}=N
<X_{p\perp}^2>_{eq}
\end{equation}
A reasonable approximation of $F_c(x)$, the Fresnel Cosine
function, is a linear function of the argument for $x\ll 1$ and
$1/2$ for $x\gg 1$.
For $p>p_{max}$,the argument of $F_c(x)$ is always
greater than $1$, and, therefore, $\tau_{p\perp}$ scales as $1/p^{5/2}$, which is a form identical to the relaxation of
unconfined modes. This result provides a self-consistency check on our calculations, since the blob model assumes that
the correlations within the blob are identical to that of an unconfined chain.

These results show that
$<X_{p\perp}^2>_{eq}$ decreases monotonically with $p$, initially as $A-Bp^{2}$, and then as $p^{5/2}$.
The slowest
transverse mode is, therefore, $p=1$, and the associated relaxation time is given by :
\begin{equation}
\tau^{\perp} \propto  L^{10/3}/(8 \pi )
\label{eqn:Xperpsmall}
\end{equation}

\paragraph{Longitudinal Modes}
For modes with $p > p_{max}$, the functional form of the
mode amplitudes is the same as that for the transverse modes:
\begin{equation}
<X_{p||}^2>_{eq} = {N^{3/2}b^2 \over
2^{3/2}p^{5/2}\pi^2}F_c\left(\sqrt{2 p \over p_{max}}
\right) \label{eqn:Xpar}
\end{equation}
For $p < p_{max}$, we use Eq.(\ref{eqn:blobscalingpar}) directly in Eq.(\ref{eqn:modeeqexact}).
In the limit of small slit widths, $L \ll N b$, we obtain a closed form expression\cite{Shen01}:
\begin{equation}
<X_{p||}^2>_{eq}\sim {(1-(-1)^p) N^2\over p^{4}L^{2/3}}+(O){L^{2/3}}\label{eqn:modeeqparsmalltube}
\end{equation}
In the small $p$ regime, therefore,  the odd and even modes behave differently.   The odd modes exhibit a $p^{4}$
decay of $<X_{p||}^2>_{eq}$, and dominate for small $L$, since the leading term in Eq. \ref{eqn:modeeqparsmalltube}
vanishes for the even modes.  For $p > N/g$,  Eq. \ref {eqn:Xpar} implies that $<X_{p||}^2>_{eq}$ decays as
$p^{5/2}$

The longest relaxation time in the longitudinal direction is determine by the relaxation of the $p=1$ mode. and is
given by:
\begin{equation}
\tau^{||} = 2N^{3}/L^{2/3}
\label{longestpar}
\end{equation}
Unlike $\tau^{\perp}$, this time-scale {\it increases} as the slit width $L$ decreases.  The equilibration time is
dominated by the relaxation of the $p=1$, longitudinal mode, and as we will show from the simulations, corresponds to
a collective rearrangement of the chain where the end-to-end vector changes sign.

In the next sections, we will use the results for the Rouse modes to analyze the dynamical properties of the confined chain.

\subsection{Collective Dynamics}
\begin{figure}[thbp]
 \centering
 \includegraphics[width=\columnwidth,clip=true]{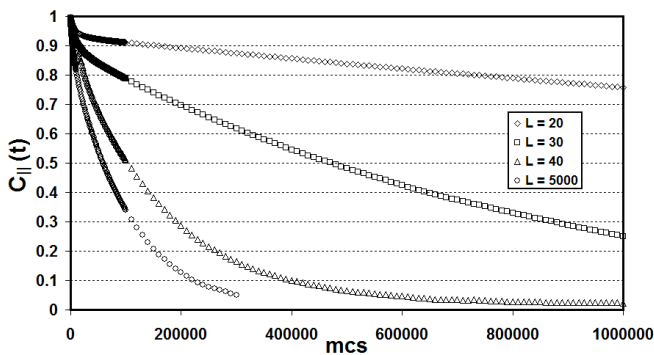}
 \caption{Simulation Results: End-to-end correlation function for component parallel to the walls for a chain with
$N=50$, $R_{g}=53$,
and for slit widths of 20, 30, 40, and 5000 (unconfined).}
 \label{cpar}
 \end{figure}

\begin{figure}[thbp]
 \centering
 \includegraphics[width=\columnwidth,clip=true]{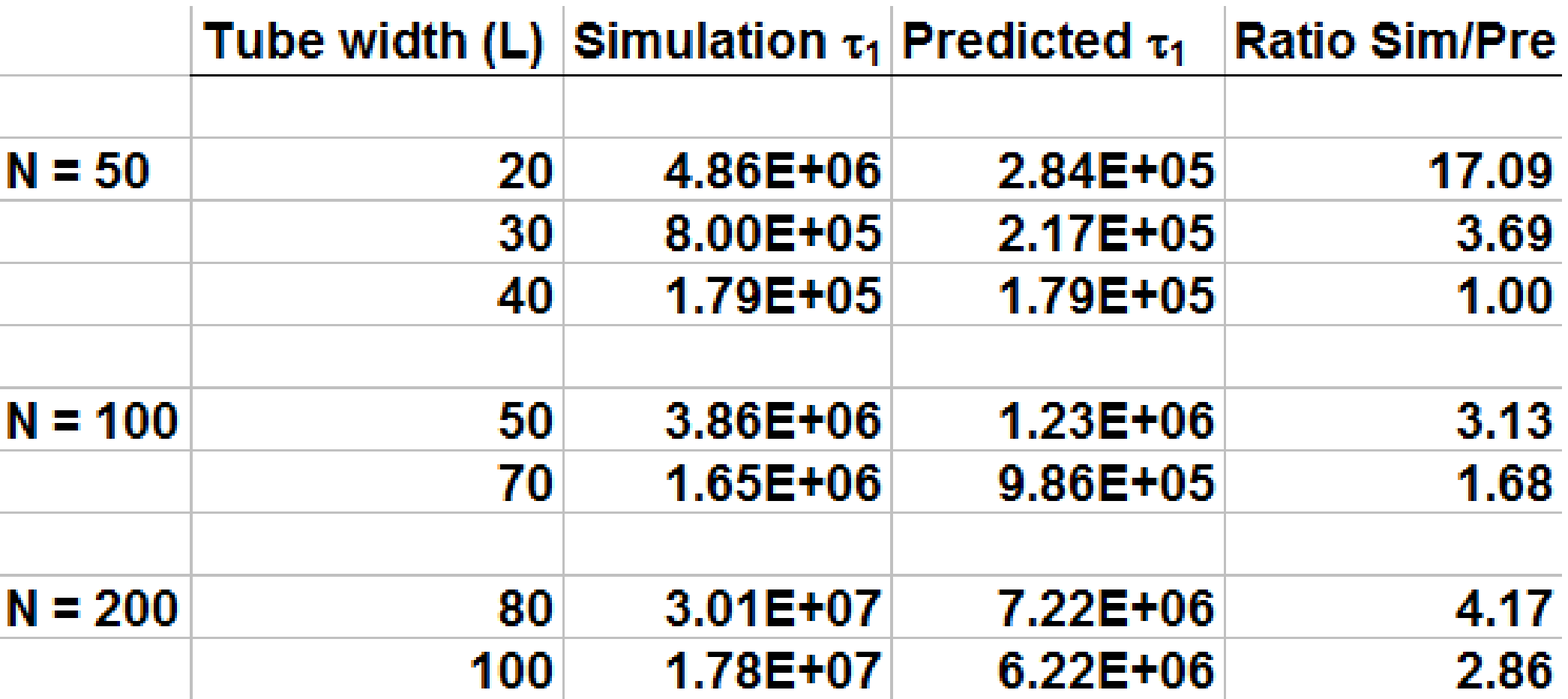}
 \caption{Simulation Results: Observed value of longest odd mode versus predicted value. A factor of 16.75 is chosen as to match up the time scales at one point $(N=50,L=40)$. This factor absorbs any
 dependence on the monomer size $b$.  The slit width is in units of lattice spacing.}
 \label{chart:expectedtau1}
 \end{figure}

The longest relaxation time of an unconfined polymer chain is associated with the relaxation of the end-to-end vector,
$\bf P(t) = \bf R_{N} (t) - \bf R_{0} (t)$
and is reflected in the autocorrelation function:
\begin{equation}
C(t) = \langle \bf P(t) \cdot \bf P(0) \rangle
\label{eq:unconCt}
\end{equation}
The Rouse model result for an unconfined,  self avoiding polymer is\cite{Doi95}:
\begin{equation}
C(t)\sim \exp\left(-{t\over \tau_{1}}\right)
\end{equation}
where $\tau_{1}$ is the relaxation time of the $p=1$, unconfined Rouse mode: $\tau_{1} \sim N^{3/2}$ in 2D\cite{Doi95}.
\begin{figure}[h]
\centering
\includegraphics[width=\columnwidth,bb=0 0 888 490,clip=true]{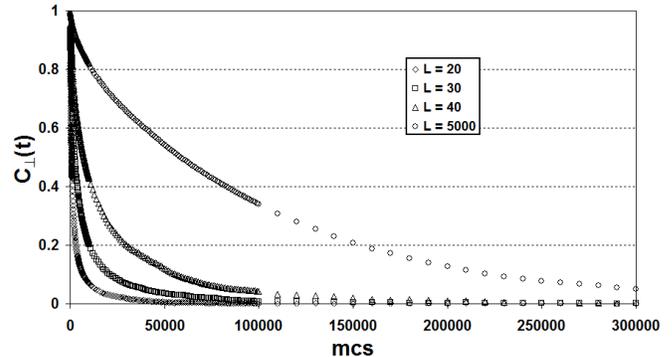}
\caption{Simulation Results: End-to-end correlation function in the perpendicular direction for a chain 50 monomers long and for slit widths of 20, 30, 40, and 5000 lattice spaces.}
\label{cperp}
\end{figure}
In a slit,
we need to look at the relaxation of the perpendicular and parallel components of $\bf P(t)$, separately.  Writing
$C_{\perp}(t)$ and $C_{||} (t))$ in terms of the Rouse modes\cite{Doi95}, it becomes evident that the long-time
relaxation is exponential with a time scale given by the longest-lived  transverse and longitudinal modes,
respectively.  From the analysis of Rouse modes for the confined polymers, it follows, therefore, that
\begin{eqnarray}
C_{\perp}(t) &\sim &\exp\left(-{t\over \tau^{\perp}} \right) = \exp \left (-2^{1/2} \pi^{2}t \over L^{10/3}\right)\nonumber\\
C_{||}(t) &\sim & \exp\left(-{t\over \tau^{||}} \right) = \exp \left (-t L^{2/3}\over 2 N^{3}\right)
\label{eq:autocorrelation}
\end{eqnarray}
These results demonstrate that the collective dynamics of a polymer confined to a
slit is qualitatively different from an unconfined polymer.
The relaxation time of the longitudinal component of the end-to-end vector {\it increases}
dramatically as $L$ is decreased, diverging as $L \rightarrow 0$.  This increase in relaxation time  has been observed in
earlier simulations\cite{Arno07} and, as shown in Figs. \ref{cpar} and \ref{chart:expectedtau1} is captured by our
simulations.
Since the time scale associated with changes in orientation of the end-to-end vector increases as
$N^{2}L^{-2/3}$, long Monte Carlo runs are required to ensure that chain configurations and
their mirror images are sampled equally.  In order to keep the simulations manageable, we studied relatively short
chains, $N=50$ monomers long.   The results were averaged over $10^9$ mcs after initial equilibration of $10^8$ mcs steps and for $5$
different initial configurations.

\begin{figure}[thbp]
 \centering
 \includegraphics[width=\columnwidth,clip=true]{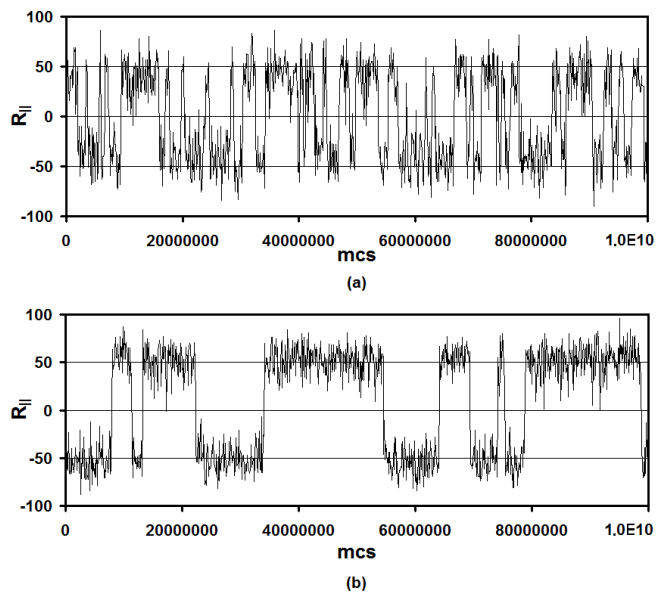}
 \caption{Simulation Results: Time trace of the longitudinal component of the end-to-end vector $\bf P(t)$ for a chain with $N=50$, $R_{g}=53$,
and for slit widths of (a) 40 and (b) 30.}
 \label{timetrace}
 \end{figure}

In the unconfined polymer, the
slowest mode corresponds to the reorientation of the end-to-end vector and the time scale increases as $R_{g}^{2}$.
Under confinement, the slowest mode still involves reorientation, but now the time scale is expected to go as
$R_{||}^{2} \sim N^{2} L^{-2/3}$, as obtained from our analysis of the Rouse modes.
Based on these arguments, the reorientation events are  expected to
become increasingly rare as $L$ is decreased.  This expectation is borne out in our simulations, as evidenced by the
time traces in Fig.\ref{timetrace}.

One way of describing fluctuations of the confined polymer is to say that
the chain fluctuates
about a fixed shape, characterized by the direction of ${\bf P}(t)$, changing shape only occasionally.  The fluctuations about the shape are described by the transverse Rouse modes, which as Eq.\ref {eqn:Xperpsmall} shows,  relax faster as the slit width is decreased, and
are independent of the size of the chain.  Results from our simulations shown in Fig. \ref{cperp}  are consistent with these predictions.
In simulations of a model polymer confined in a 2D box\cite{Rahm07},
it has been observed that the slow modes correspond to shape changes, and that the slow dynamics is not reflected
directly in the monomer motion.    The shape of the polymer is a simpler variable in the slit geometry, but a distinct
separation between time scales associated with shape changes and time scales associated with individual monomer motion
is predicted by the generalized Rouse dynamics, and seen in the simulations.

In the next section, we analyze the motion of monomers using the framework of the Rouse modes.

\subsection{Monomer Motion}In this section, the effective Rouse modes are used to make detailed predictions  for the mean-squared-displacement (MSD) of monomers.
The theory for the  MSD of monomers should provide a framework for understanding the dynamics inside biological
cells where motion has been observed to be subdiffusive\cite{Ther06}.

Starting from Eq.(\ref{eqn:modeeq}) and making use of the fact the perpendicular mode index has a lower bound of $N/g$, the number of blobs,
it can be shown that the displacement squared of a monomer in the direction perpendicular to the walls is
\begin{equation}
<(x_{n\perp}(t)-x_{n\perp}(0))^2> = \sum_{p=(N/g)}^{\infty}{4\over k_{p\perp}}\left[1-exp\left(-{t\over \tau_{p\perp}}\right)\right]
\label{eqn:perpmot}
\end{equation}
Likewise the expression for the parallel monomer motion is
\begin{equation}
<(x_{n||}(t)-x_{n||}(0))^2>={2t\over N}+\sum_{p=1}^{\infty}{4\over k_{p||}}\left[1-exp\left(-{t\over
\tau_{p||}}\right)\right]\label{eqn:parmot}
\end{equation}
\begin{figure}[h]
 \centering
 \includegraphics[width=\columnwidth,clip=true]{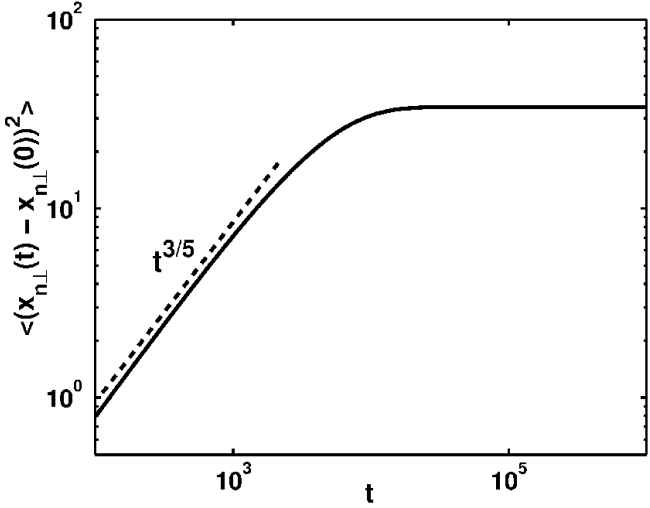}
 \caption{Prediction for perpendicular motion from theory for a chain of 200 monomers, monomer size 2.8, and slit width
30 ($L/R_{g}=1/5$). (see Eq.(\ref{eqn:perpmot}))}
 \label{perpmottheory}
 \end{figure}

\begin{figure}[h]
\centering
\includegraphics[width=\columnwidth,clip=true]{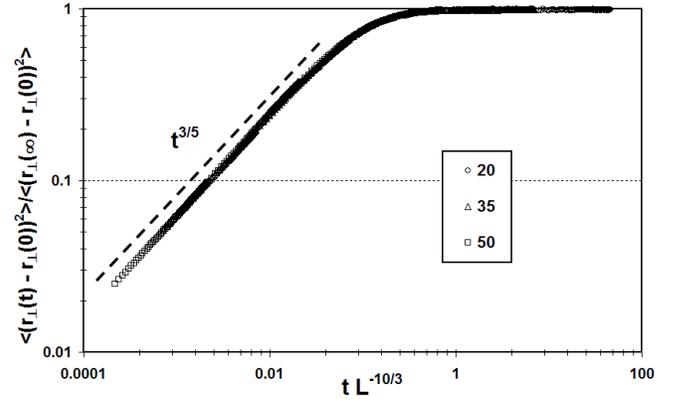}
\caption{Simulation Results: Plots of MSD of the perpendicular component of the middle monomer in a chain with $N=200$
The MSD have been rescaled by the maximum value of the saturation value at each $L$, and time has been scaled by, $L^{-10/3}$, to capture the scaling of
$\tau^{\perp}$. These scalings lead to data collapse for $L= 20~, 35~, {\and}
50$.}
\label{perpmonmotscaled}
\end{figure}

\begin{figure}[h]
\centering
\includegraphics[width=\columnwidth,clip=true]{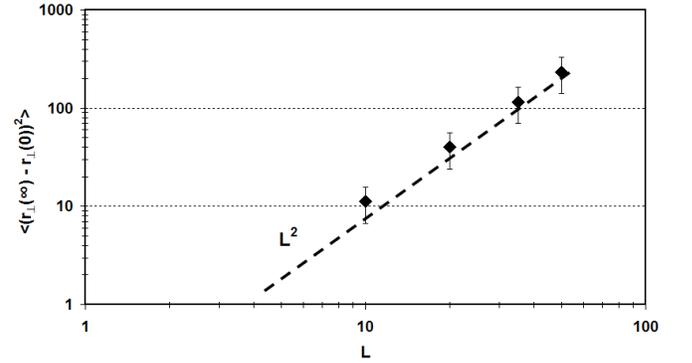}
\caption{Simulation Results: Plots of saturation value of displacement squared of the perpendicular component of the middle monomer in the chain recovered from simulations. The chain is 200 monomers long and plots are for slit widths of 10, 20, 35, 50.}
\label{perpmotsat}
\end{figure}

The main features of the perpendicular motion, as predicted by
(\ref{eqn:perpmot}), are depicted in Fig.(\ref{perpmottheory}).   Since  $k_{p \perp}$ has the same functional
form as the unconfined polymer modes for $p \geq N/g$, the MSD exhibits a $t^{3/5}$ power law that saturates to a value
proportional to $L^{2}$ for $t \gg \tau^{\perp}$, the relaxation time of the $p=N/g$ mode.  The explicit exponent of
$3/5$ can be obtained by analyzing Eq. \ref{eqn:perpmot} for $t \ll \tau^{\perp}$.   In this limit, the sums can be
replaced by integrals, and we obtain:
\begin{equation}
<(x_{n\perp}(t)-x_{n\perp}(0))^2> \approx{2^{16/5}b^{4/3} \over 3\pi^{4/5}} t^{3/5} \Gamma_{q}(2/5) ~.\label{eqn:perpmotapprox}
\end{equation}
In the above equation, $\Gamma_{q}$ is the Gamma function, and $q={b^{4/3} \pi^2 t\over 2^{1/2}L^{10/3}}$. The results
of simulations for a $N=200$ chain confined to slit widths ranging from $20$ to $50$ are shown in Fig.
\ref{perpmonmotscaled}.  As predicted by the theory, MSD for different values of $L$ can be collapsed on to each other by scaling the
time $t$ by $\tau^{\perp} \sim L^{10/3}$ and scaling the MSD by its maximum value which increases as $L^{2}$ (Fig.
\ref {perpmotsat}).

The MSD in the longitudinal direction, described by Eq. \ref {eqn:parmot}, is expected to exhibit multiple regimes.
The functional form of of the effective spring constants, $k_{p ||}$,  changes at $p=N/g$ with a $p^{4}$ increase at
low $p$ (for the odd modes) and a $p^{5/2}$ increase for the high $p$ modes.   The analysis is further complicated by
the difference between odd and even modes for low mode numbers.
\begin{figure}[h]
 \centering
 \includegraphics[width=\columnwidth,clip=true]{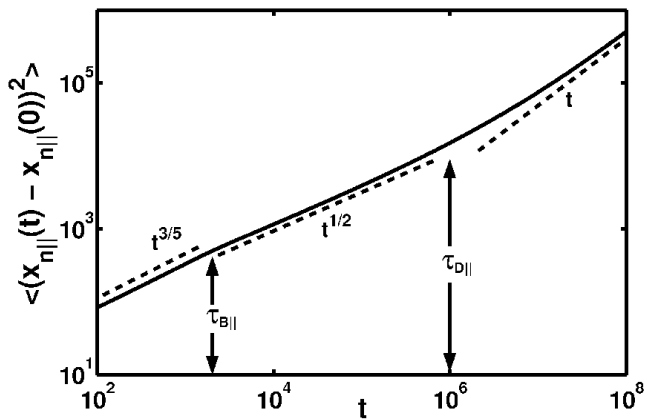}
 \caption{Results  for the MSD in the longitudinal direction
obtained by numerical evaluation of
Eq.(\ref{eqn:parmot}) for a chain of 200 monomers, monomer size 2.8, and slit width 30.  The dashed lines and the
arrows mark the power laws and the crossover times predicted by the short-time approximation, Eq. (\ref{eqn:parmotapprox})}
 \label{parmottheory}
 \end{figure}
Even though there is no closed form expression for the MSD in the longitudinal direction, we can recover useful information about
the functional forms and crossover times. To accomplish this, we
look at  times much less than the longest relaxation time
of the system. With this assumption, the sum in
(\ref{eqn:parmot}) can be approximated as an
integral and  results in the following closed form expressions\cite{Kalb}:
\begin{eqnarray}
<(x_{n||}(t)-x_{n||}(0))^2> \approx {2 t\over N}+\nonumber\\
{2^{16/5}b^{4/3} \over 3\pi^{4/5}} t^{3/5}
\Gamma_{b^{4/3}\pi^2 t\over 2^{1/2} L^{10/3}}(2/5)+\\
{2^{7/2} b^{2/3} L^{1/3} \over \pi^{1/2}} t^{1/2} Erf\left({\pi
 b^{2/3} t^{1/2} \over 2^{1/2} L^{5/3}}\right)\nonumber
\label{eqn:parmotapprox}
\end{eqnarray}

where we have made explicit use of Eq.  (\ref{eqn:Xpar}).
As expected, the ${0}^{th}$ mode leads to purely diffusive motion.  The system is dominated by the $t^{3/5}$ behavior
until a crossover time of  associated with relaxations within the blob, $\tau_{B||}$:
\begin{equation}
\tau_{B||}={2 L^{10/3} \over \pi^2 b^{4/3}} ~,
\label{eqn:tauBpar}
\end{equation}
which is proportional to $\tau^{\perp}$.   For times longer than $\tau_{B||}$, $t^{1/2}$ is the dominant
behavior, and this subdiffusive motion becomes diffusive as the $0{th}$
mode starts dominating at $\tau_{D||}$\cite{Genn71}:
\begin{equation}
\tau_{D||}={8 b^{4/3} N^2 L^{2/3}\over \pi}
\label{eqn:tauDpar}
\end{equation}
These time scales, and the power laws are shown in Fig.\ref{parmottheory}.

The different power-law regimes correspond to physically different relaxation mechanisms.  At the shortest time
scales, the relaxation is not affected by the confinement, and the MSD follows that of the unconfined, self-avoiding
chain.  At intermediate times, the polymers can be thought of as an ideal chain of blobs, and blob compression
dominates the dynamics\cite{Broc77}.   The MSD, therefore, shows the sub-diffusive behavior characteristic of
an ideal chain with no excluded volume constraint.  The long-time diffusive motion reflects the center-of-mass motion.
An  interesting observation to be made is that the diffusive behavior sets in at earlier and earlier times as the slit
width $L$ is decreased.  This apparently counterintuitive result, however, follows from the fact that with increasing
confinement, transverse fluctuations decay faster, ($\tau^{\perp} \sim L^{10/3}$), and the longitudinal motion of the
polymer essentially looks like that of a rigid rod.  The reorientation events, which are the slowest dynamical process
in the system, do not affect the MSD of monomers.

\begin{figure}[h]
\centering
\includegraphics[width=\columnwidth,clip=true]{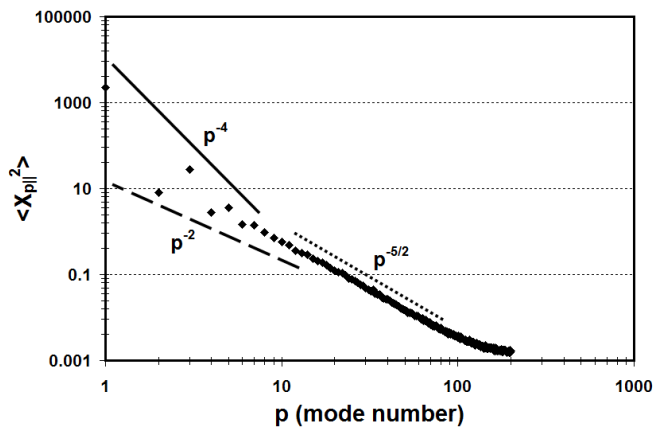}
\caption{Simulation Results: Mode excitation vs mode number $p$ extracted from simulation for a chain 200 monomers long and a slit width of 30 lattice space. The saturation value for the largest modes is due to the chain being discrete instead of continuous. Solid and dashed lines represent the linearized Rouse theory predictions.}
\label{modesparsim}
\end{figure}

The above analysis relies on our deductions for the functional form of the amplitudes $\langle X_{p ||}^{2} \rangle$,
which determine $k_{p ||}$.   We have extracted the Rouse mode amplitudes from the simulations (see
Fig.\ref{modesparsim}) to directly verify these functional forms.  We find that for mode numbers $p$ greater than a
characteristic value $p_{s}$, the amplitudes decay as $p^{-5/2}$, just as they do in the unconfined chain. However, for
$p \leq p_{s}$, there is an observed splitting in the even and odd modes, with the
even modes seemingly exhibiting a $p^{-2}$ behavior and the odd modes decaying as $p^{-4}$.  Since we can obtain data
only over a limited range (determined by the size of the chain), these power laws are only approximate.  The crossover
at $p_{s}$ is a robust phenomenon.  From our theoretical analysis, we expect $p_{s} = N/g$.  For a chain of $N=200$,
$b=2.8$ (characteristic of the bond-fluctuation model), and a slit width of $L=30$, the blob model predicts
that $N/g = 8.0$.  The simulation data (Fig.\ref{modesparsim})) is in surprisingly good agreement with this prediction.
Our simulation results, therefore, support the conclusion that the statistical properties of the
confined polymer can be understood in terms of the effective Rouse modes obtained from our
theoretical framework.

\begin{figure}[h]
 \centering
 \includegraphics[width=\columnwidth,clip=true]{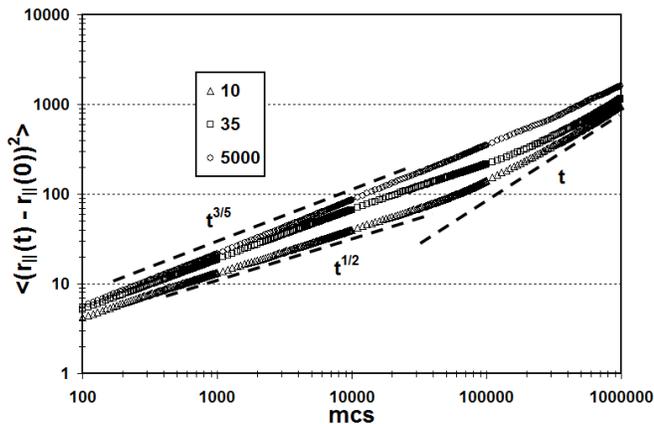}
 \caption{Simulation Results: MSD of the middle monomer parallel to the slit walls for a chain of length 200 and the slit
widths shown in the figure. For the unconfined chain, $L=5000$, the diffusive regime is barely visible at $t= 10^{6}$
mcs.  In contrast, for $L=10$, the diffusive regime is visible at $t=10^{5}$ mcs.}
 \label{monmotpar}
 \end{figure}

\begin{figure}[h]
 \centering
 \includegraphics[width=\columnwidth,clip=true]{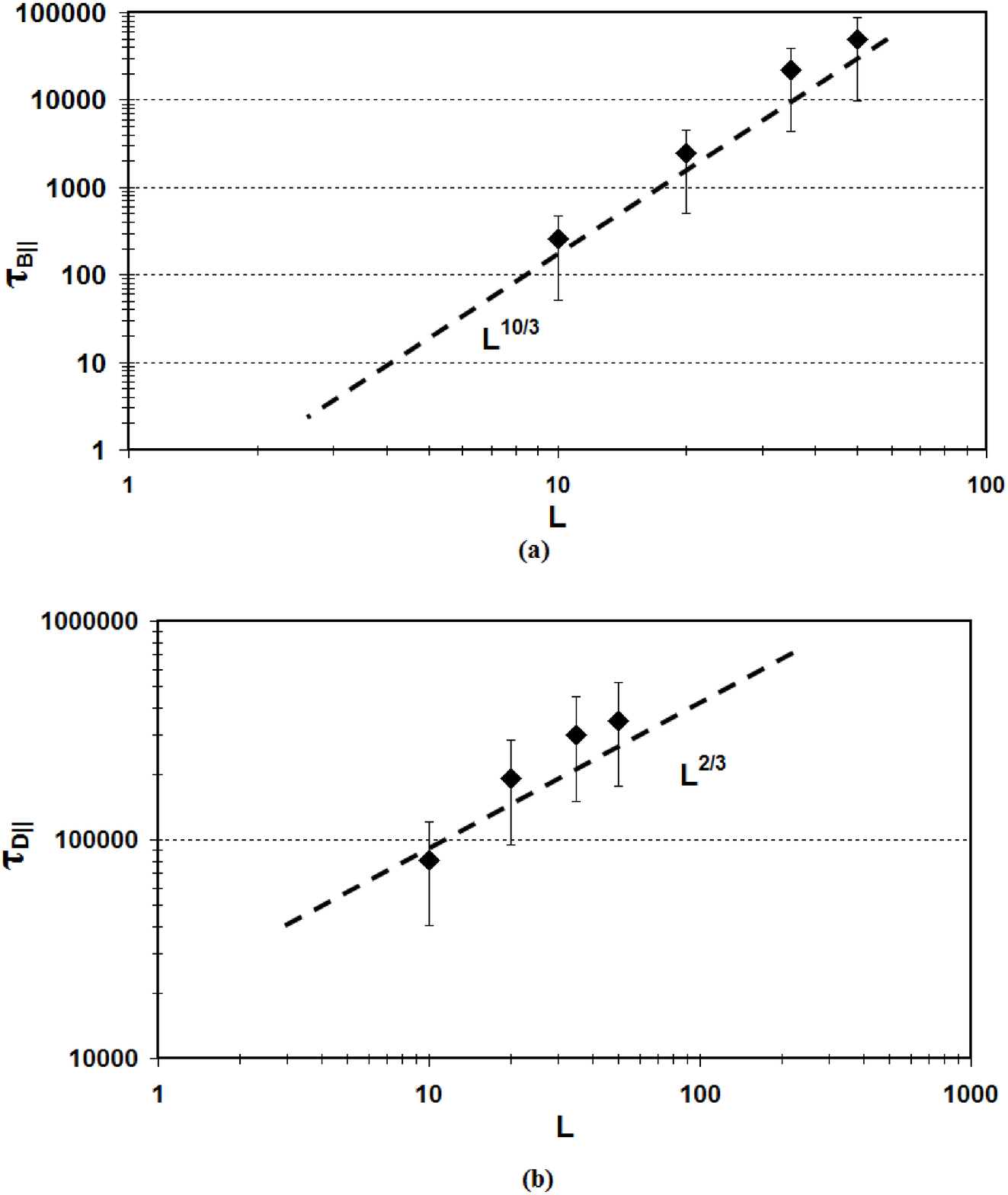}
 \caption{Simulation Results: $\bf{(a)}$ Plot of first crossover time from $t^{3/5}$ to $t^{1/2}$ ($\tau_{B||}$) from Fig.\ref{monmotpar}.$\bf{(b)}$ Plot of first crossover time from $t^{1/2}$ to $t$ ($\tau_{D||}$) from Fig.\ref{monmotpar}.}
 \label{tbtd}
 \end{figure}

We have used our simulations to  carefully measure the MSD of monomers in the longitudinal direction, and the results are shown in
Figs. \ref{monmotpar} and \ref{tbtd}.   Our results are consistent with the theoretical
predictions. The short time behavior is governed by a sub-diffusive, $t^{3/5}$ regime, associated with the unconfined
polymer. There is a hint of an intermediate regime, governed by the slower power law of $t^{1/2}$. Finally, at long
times, the MSD is observed to increase  linearly with $t$. The two characteristic times associated with the crossover
from
$t^{3/5}$ to $t^{1/2}$ ($\tau_{B||}$) and $t^{1/2}$ to $t$ ($\tau_{D||}$) are plotted in Fig.\ref{tbtd} and both times
are observed to be monotonically decreasing with the tube widths, approximately obeying the scaling predicted by Eqs. (\ref{eqn:tauBpar}) and (\ref{eqn:tauDpar}).

\section{Conclusion}

We have presented a theoretical framework for analyzing the dynamics of confined, self-avoiding polymers that is valid
for any confining geometry.
The formalism is based on renormalized, effective, Rouse modes.   Through extensive simulations, we have shown that
he predictions of the theory are borne out for a polymer chain confined to a slit.

Given the Rouse mode framework, both collective dynamics and single monomer motion can be determined
from a knowledge of the mode relaxations.  Our analysis shows that confinement introduces multiple time scales
into the dynamics.   The slowest  mode is associated with the reorientation of the end-to-end vector, and its
relaxation time increases as $L^{-2/3}$ with decreasing slit width.  Transverse fluctuations of the chain relax on a time
scale that decreases as $L^{10/3}$.  This latter time scale shows up in the monomer dynamics as a crossover time from
a $t^{3/5}$ subdiffusive behavior, characteristic of the self-avoiding polymer, to a $t^{1/2}$ subdiffusive behavior,
characteristic of an ideal, phantom chain.   In an unconfined polymer this intermediate regime is absent.   Under
confinement, the monomers exhibit pure diffusion in the longitudinal direction at times longer than $\tau_{D ||} \sim
L^{2/3}$, implying that the diffusive motion sets in at earlier times as the slit width is decreased.
This observation can be understood by noting that a SAW behaves increasingly as a rigid rod as the slit width
decreases and the dominant motion is that of diffusion of the center of mass.   It should be remarked that in an
unconfined chain the monomer motion becomes diffusive at time scales long compared to the reorientation, or rotational
time scale.  For the confined polymer, the
single monomer motion does not
exhibit the slowing down of the dynamics associated with reorientation events in which the chain undergoes a complete
back bending, reversing direction. The Rouse dynamics of an unconfined polymer is characterized by a single time scale,
$\tau_{1} \sim N^{2\nu}$.  As we have seen, the dynamics of a polymer confined to a slit has three distinct time scales,
which scale with the slit width as $L^{10/3}$, $L^{2/3}$, and $L^{-2/3}$.  The separation between the shortest, $\sim L^{10/3}$ and
the longest, $L^{-2/3}$ time scales grows as $L$ is decreased, resulting in a clear separation of the longitudinal
and transverse dynamics.

There are small discrepancies between theory and simulations that can be ascribed to
one or both of two effects: (a) the lattice spacing associated
with the simulation model, (b) the finite length of the polymer, and the
limitations of statistics.
The one significant difference between theory and simulations is in the $L$-dependence of the longest relaxation time,
$\tau^{||}$.   Further studies are needed to investigate the origin of this discrepancy.
Within the framework of Rouse modes, a more sophisticated framework can be constructed by considering a
renormalization of the single-monomer friction coefficient, and its predictions compared to simulations.  Adhering to
the Rouse mode framework preserves the connection between static correlations and dynamics that is extremely desirable
for parsing the complex dynamics of polymers under confinement.

The results presented in this paper demonstrate that the difference in dynamics in the longitudinal and transverse directions
provides detailed information about the effect of tube confinement on a self-avoiding polymer.  These results apply as long as the confinement is weak, {\it i.e.},  the longitudinal dimension is larger than the
unconfined radius of gyration of the polymer.   If the longitudinal dimension is comparable or smaller than the radius of gyration, then our results will be altered but
the anisotropy, and especially the separation of time scales will persist and be prominent for aspect ratios larger than unity.
The latter situation applies to bacterial chromosomes.   Recent research has focused on studying the role played by polymer confinement
on chromosome dynamics and segregation\cite{Suck07}.
Our results indicate that resolving the transverse and longitudinal components in measurements of chromosomal dynamics would provide important diagnostic tools.
Such measurements will aid the understanding of the effects of cell geometry and
crowding on the fine scale dynamics of bacterial chromosomes\cite{Ther06}, and the relevance of intrinsic polymer dynamics to biological processes.

We would like to thank  Jeremy Schmit and Jan\'{e}
Kondev for many insightful discussions, and Michael Hagan for a careful reading of the manuscript. This work was
supported in part by NSF-DMR 0403997.
\bibliography{sourcesa}

\end{document}